\documentclass[12pt,preprint]{aastex}
\shorttitle{Jet Activity in 3C 273}
\shortauthors{Stawarz}
\usepackage{natbib}

\begin{document}

\title{On the Jet Activity in 3C 273 }

\author{\L ukasz Stawarz}
\affil{
Obserwatorium Astronomiczne, Uniwersytet Jagiello\'nski,\\
ul. Orla 171, 30-244 Krak\'{o}w, Poland}
\email{stawarz@oa.uj.edu.pl}

\begin{abstract}

In this paper we comment on the possibility for intermittent jet activity in quasar 3C 273 on different time-scales. We propose, that striking morphology of the large-scale radio jet in this source, as well as the apparent lack of its counterpart on the opposite side of the active center, may be explained in a framework of a restarting jet model. In particular, we propose that 3C 273 radio source is intrinsically two-sided, and represents an analogue of double-double radio galaxies, but only inclined at a small angle to the line of sight. In this case, the apparent one-sideness of the kiloparsec-scale radio structure may be due to combined Doppler and time-travel effects alone, if the 3C 273 large-scale jet itself is relativistic and matter-dominated. We also propose, that knotty morphology of the discussed jet, which is observed now additionally at optical and X-ray frequencies, indicates modulation in the jet kinetic power. This, together with the variability of the jet at small (parsec) scales, indicates that the jet activity in 3C 273, and possibly in other similar sources, is variable/modulated/intermittent over many different time-scales.

\end{abstract}

\keywords{galaxies: active---galaxies: jets---quasars: individual (3C 273)}

\section{Introduction}

3C 273 is the well-known quasar at the redshift $z = 0.158$, discovered together with its famous large-scale jet at the very beginning of quasar optical researches \citep[see a review by][]{cou98}. At radio frequencies it appears as the one-sided, core-dominated source of non-thermal radiation, with the radio jet extended up to a few tens of kpc from the core, similarly to its optical counterpart. The jet is also prominent at X-ray frequencies, although the origin of this radiation is still being debated (in contrast with the radio-to-optical emission, whose synchrotron nature is surely established). Until now, neither the jet composition nor its velocity on the kpc-scales is known. In addition, the unusual morphology of the radio structure -- lack of an extended radio cocoon in particular -- is not easy to interpret, suggesting, in the opinion of some authors like \citet{cla86}, Poynting flux-dominated flow \citep[see in this context also][]{lin89,kom99}. Finally, the existence of the counterjet in 3C 273 is also a controversial issue, because the jet-counterjet brightness asymmetry at the jet terminal point is exceptionally high, with a lower limit $10000:1$ \citep{dav85,con94}. At the working surface of the jet only subrelativistic bulk velocities, and therefore minor Doppler-hiding effects, are expected \citep[e.g.,][see also section 2.3 below]{beg84}. Hence, some authors \citep{con81,dav85} have concluded that 3C 273 is an intrinsically one-sided jet source, that seems to be consistent with the lack of \emph{any} radio feature (like some extended lobe, for example) on the 3C 273 counter-side.

In this paper we comment on the possibility for intermittent jet activity in quasar 3C 273 on different time-scales. We propose that the striking morphology of the large-scale jet in this source, as well as the apparent lack of its counterpart on the opposite side of the active center, may be explained in a framework of restarting and modulated jet model. In particular we propose that 3C 273 radio source is intrinsically two-sided, and represents an analogue of double-double radio galaxies \citep{sch00a}, but only inclined at small angle to the line of sight, and in addition that the 3C 273 large-scale jet itself represents a continuous flow which is relativistic, matter-dominated and highly modulated in kinetic power. We discuss how present observations support this idea, and how the future ones can eventually verify it. Section 2 contains a brief presentation of the 3C 273 jet. In Section 3 we consider the proposed model of intermittent jet activity in this source. Discussion and final conclusions are presented in the last Section 4.

\section{The jet in 3C 273}

\subsection{Small-scale jet}

VLBI observations of the 3C 273 \citep[e.g.,][]{pea81,unw85,bir85,coh87,zen88,zen90,kri90,abr96,abr99} show a small-scale radio jet composed from several components (blobs), emanating every few years from the flat-spectrum unresolved radio core at the position of the quasar. It was suggested that the curved trajectories and variability pattern of these blobs may be explained by helical jet structure. The jet components located within the inner few tens of mas from the active center are characterized by apparent superluminal velocities, which imply a small viewing angle of the pc-scale jet, $\theta \sim 10^{\circ}$, and its high bulk Lorentz factor, $\Gamma_{\rm jet} \geq 5$. On the other hand, the inclination of the nuclear jet is not expected to be smaller than $\sim 10^{\circ}$, because 3C 273 does not belong to the (regular) blazar class, due to the presence of a thermal UV bump in its spectrum. The orientation of the large-scale jet in the plane of the sky differs from the orientation of the inner jet by about $\sim 20^{\circ}$, which is consistent with the apparent bend of the flow occurring at $\sim 10$ mas from the core.

\subsection{Large-scale jet}

3C 273, as observed at radio frequencies \citep[e.g.,][]{con81,fla85,fol85,dav85,con93}, reveals a large-scale jet with the non-thermal radio continuum characterized by the spectral index $\alpha_{\rm R} \sim 0.8$, which changes very smoothly along the jet. The radio jet extends continuously from the core up to $\sim 23''$ away, being composed from several brightenings (knots) linked by the intense diffuse emission. Starting from the first prominent knot placed $\sim 13''$ from the quasar and labeled as `A', the radio jet is approximately $\sim 1''$ wide. Its brightness increases with the distance and peaks at the jet terminal point $\sim 20 - 23''$ from the core, in the so-called `head' of the jet. The head itself, labeled as `H', reveals the triple sub-structure (H3, H2 and H1 going along the jet axis, respectively). At the southern side of the jet, an extended low-brightness region is present, being characterized by the very steep spectrum with $\alpha_{\rm R} \sim 1.5$. While the jet head is interpreted as the hot-spot region, the southern extension of the jet is widely considered as a manifestation of the fragmentary or just strangely shaped radio lobe (see sections 2.3 and 2.4 below). The radio jet emission is polarized at the level $10 - 20\%$. The inferred magnetic field, with equipartition intensity $\sim 10^{-5} - 10^{-4}$ G, is parallel to the jet axis up to $20.5''$ from the core (H3 sub-component of the jet head), where it suddenly becomes transverse. This change in the magnetic field structure is consistent with the interpretation of the H2 region as the hot-spot shock \citep[e.g.,][]{mei86}.

Optical observations of the 3C 273 large-scale jet \citep[see, e.g.,][]{roe91,bah95,roe96,neu97,jes01,jes02} reveal its non-thermal emission at infrared-to-ultraviolet frequencies, pronounced from the region placed between the knot A and the jet head H, i.e. from $\sim 12''$ up to almost $\sim 22''$ from the core. The optical jet is narrower than its radio counterpart, with the $\sim 0.7''$ total width being approximately constant along the flow. It is composed from several knots coinciding with the radio ones, linked by the diffuse inter-knot emission, and inclined to the jet axis at about $\sim 45^{\circ}$. The nature of these knots is not certain. It was proposed, for example, that they may represent oblique shocks formed due to magneto-hydrodynamical instabilities within the propagating flow, or that they may be manifestations of a helical magnetic field configuration on large scales. The optical emission is the strongest at the position of the first knot A, and only very weak at the jet terminal point H. Comparison of the radio and optical morphologies therefore strengthens the conclusion that the jet head is physically distinct from the rest of the large-scale jet, and indicates in addition a conspicuous radial structure of the jet itself. In particular, the jet seems to consist from the proper cylindrical beam distinct at optical frequencies, which is surrounded by the diffuse `cocoon' dominating radio emission. The optical radiation is polarized at the level $5 - 20 \%$, which indicates, with little doubt, its synchrotron nature. The changes in the level of polarization and the inferred magnetic field configuration are similar to those observed at radio frequencies, although some minor differences are noted. The radio-to-infrared power-law slope of the 3C 273 large-scale jet, $\alpha_{\rm R-IR} \sim 1$, is remarkably constant along the flow. On the other hand, the optical spectral index declines very smoothly from $\alpha_{\rm O} \sim 0.5$ in the knot A region, up to $\sim 1.5$ at the jet terminal point. Interestingly, at some jet regions spectral hardenings at ultraviolet frequencies over the extrapolated radio-to-optical continua are observed \citep{jes02}. The overall spectral behavior of the optical jet in 3C 273 strongly suggests in-situ particle acceleration acting within the whole jet body \citep{jes01}.

The 3C 273 large-scale jet is also known for its X-ray emission \citep{har87,roe00,mar01,sam01}. This emission occurs first in the knot A region, thereby being the strongest, and then decreases systematically to reach the background level before the main sub-component of the jet head, H2, where the jet radio emission peaks. Hence, the main difference between the X-ray jet and its radio and optical counterparts is that the X-ray brightness generally declines along the flow, while the optical one is roughly constant (apart from the knotty structure) and the radio one increases significantly. The featureless $1 - 10$ keV continuum of the 3C 273 jet is widely believed to be non-thermal in origin. The X-ray emission of the knot A, which constitutes $\sim 40\%$ of the total X-ray jet luminosity, is characterized by the relatively flat X-ray spectral index, $\alpha_{\rm X} \sim 0.6$. The X-ray and the optical fluxes for this region are both slightly above the extrapolation of the radio-to-infrared continuum, suggesting therefore strongly the synchrotron origin of the observed keV photons, but on the other hand excluding the possibility of modeling the radio-to-X-ray radiation as a synchrotron emission of a single, one-power-law electron energy distribution \citep{jes02}. For the other knots the situation is even less clear, as the X-ray emission thereby exceeds more significantly the continua extrapolated from the lower frequencies. It was proposed, that a possible explanation for the 3C 273 jet X-ray radiation could be comptonization of the CMB photon field \emph{if} the jet emission on large distances from the core is highly beamed, and therefore if both the jet inclination and bulk velocity are similar for the pc- and tens-of-kpc-scales. In the framework of this model the jet magnetic field is roughly consistent with the equipartition value, but on the other hand some observed spectral and morphological characteristics are hard to understand \citep[see][]{ato04,sta04}.

\subsection{Hot-spot region}

The head of the 3C 273 large-scale jet, i.e. its H component, is widely believed to be a manifestation of the hot-spot region, and in particular its brightest H2 sub-component is usually identified with the jet terminal shock, `the Mach disc' \citep{mei86,con94}. The forward speed of the respective terminal structures in most of the classical double sources is estimated from, e.g., ram-pressure arguments to be only sub- or eventually mildly-relativistic. If this is also the case for the 3C 273 jet, then the very high brightness asymmetry between the H2 region and its hypothetical counterpart on the other side of the quasar ($\sim 10^4$) cannot be explained by the Doppler effects alone. For this reason, the 3C 273 was suggested to be most likely an intrinsically one-sided source. However, the density of the ambient medium in vicinity of the 3C 273 jet -- parameter which is crucial in applying the ram-pressure arguments -- is unknown, and therefore the need for the intrinsic one-sideness of the 3C 273 radio quasar should be considered with caution.

\citet{con81} argued that the density of the thermal ambient medium surrounding 3C 273 large-scale jet, $n_{\rm amb}$, has to lie between the mean density of the Universe, assumed by them to be $\sim 2 \times 10^{-5}$ cm$^{-3}$, and the value set by the upper limit inferred from the X-ray emission of the thermal gas around 3C 273, given by \citeauthor{con81} as $\sim 5 \times 10^{-3}$ cm$^{-3}$. Hence, the forward velocity of the jet head should be $0.02 < \beta_{\rm head} < 0.3$ for the large-scale jet viewing angle $\theta \sim 10^{\circ}$ anticipated from the pc-scale jet observations. \citet{con81} noted that smaller values of $n_{\rm amb}$ are not expected for at least three reasons. First, the typical density of the intergalactic medium in the groups of galaxies, like the poor group to which 3C 273 host galaxy belongs, lie usually between the mean density of the Universe and the typical density found in the clusters of galaxies, $10^{-4} - 10^{-2}$ cm$^{-3}$. Secondly, the formation of the massive quasars like 3C 273 is not expected in the regions of unusually low density. And finally, excellent collimation of the 3C 273 large-scale jet (opening angle $< 2^{\circ}$) indicates that it cannot be in free expansion but instead has to be confined sideways, most probably -- according to \citet{con81} -- by the thermal ambient medium with relatively high density. \citet{con93} discussed an alternative way to estimate the density of the ambient thermal medium around 3C 273, provided by the Faraday rotation measurements, which yield ${\rm RM} \sim 2 \pm 2$ rad m$^{-2}$ everywhere in the 3C 273 radio source. Assuming that the ambient gas around the discussed object is characterized by the conditions typically found in clusters, i.e. by the ratio of thermal to magnetic field pressures being several hundred, the core radius of the hot gas with temperature $\sim 10^7$ K being $\sim 0.4$ Mpc, and by the random intergalactic magnetic field being turbulent with a reversal scale length $\sim 10$ kpc, \citet{con93} estimated the ambient gas density $n_{\rm amb} \sim 3 \times 10^{-4}$ cm$^{-3}$, in agreement with conclusions by \citet{con81}. One can note that ROSAT observations of 3C 273 \citep{roe00} revealed the presence of slightly extended thermal emission with the total X-ray luminosity $\sim 3.4 \times 10^{43}$ erg s$^{-1}$, implying central density of the hot gaseous atmosphere of 3C 273 being $\sim 6 \times 10^{-2}$ cm$^{-3}$, i.e. ten times larger from the value quoted by \citet{con81} as an upper limit.

The low value of $\beta_{\rm head}$ was adopted in a discussion by \citet{mei86} regarding physical conditions within the terminal shock of the 3C 273 jet \citep[see also][]{mei89,mei97,con94}. In modeling radio-to-optical emission of the H2 region (identified with the Mach disc) the authors obtained consistency with the model of first-order Fermi particle acceleration at non-relativistic shock front, for the large-scale jet viewing angle $\theta \sim 40^{\circ}$ and the hot-spot magnetic field (inferred from the ageing features of the synchrotron continuum) $\sim 4 \times 10^{-4}$ G in agreement with the minimum power estimates. The discrepancy between the large-scale jet inclination and the value inferred at the smaller scales from superluminal motions was ascribed to the jet bending occurring at $\sim 10$ mas from the core. The large jet inclination at kpc-scales was also claimed to be consistent with the polarization structure of the jet head. As mentioned in the previous section, observations imply the jet magnetic field as being parallel to the jet axis up to the H3 component, beyond which it becomes suddenly transverse, reaching the linear polarization of $\sim 23 \%$ in the H2 region. By interpreting this change as due to compression of the jet magnetic field at the terminal shock, one gets the required angle between normal to the compression plane and the line-of-sight $\sim 40^{\circ}$ \citep{fla85}. This interpretation then requires large bending of the jet between pc- and kpc-scales. Alternatively, relativistic aberration effect could decrease the required angle, as proposed by \citet{fla85}, who rejected the possibility for the significant bending of the 3C 273 jet, arguing that the apparent $20^{\circ}$-bend at mas scale, if corrected for the foreshortening effects, cannot be in reality greater than $\sim 5^{\circ}$. This would allow for a relativistic velocity of the H2 region, with the Doppler factor $\sim 3 - 4$. In such a case, H2 would represent rather the knot within the 3C 273 large-scale jet -- in the `standard' picture -- and not the regular terminal shock. However, other authors anticipated the large inclination of the 3C 273 kpc-scale jet on the ground of the large bending hypothesis. In particular, \citet{mei86,mei89,mei97} and \citet{con94} adopted the non-relativistic velocity for the jet head (recalling the ram-pressure arguments as presented in the previous paragraph) together with particular models explaining the observed changes in polarization within the head region \citep{lai80,caw88}. In this way they obtained quantitative agreement between the H2 synchrotron emission and the standard terminal shock model. We believe, however, that because of complex and hardly known morphology of the jet terminal regions \citep[see in this context, e.g.,][]{miz04}, and of the jet as well, one can still consider values for some parameters of the 3C 273 hot-spot being different from the ones obtained in the framework of such a one-dimensional model.

\subsection{Extended radio structure}

It is not clear if the radio cocoon surrounding the optical (and X-ray) large-scale jet in 3C 273 represents back-flowing material turned back from the working surface of the jet, or only a slower-moving (or even stationary) extension of the jet boundary layer. In the former case the ratio of its length ($\sim 10''$) to its width ($\sim 1''$) would be exceptionally high as compared to the other powerful radio sources \citep[i.e., FR II radio galaxies and radio-loud quasars, for which typical axis ratio of the radio lobes is $\sim 3$; e.g.,][]{beg89,kai97}. Such a narrow-lobe morphology cannot be reproduced in numerical simulations in case of a matter-dominated, low-density jet. On the southern side of the jet, the cocoon joins smoothly with the diffuse radio region, believed to be the proper radio lobe \citep{dav85,con93}. Until now, no similar structure had been found either on the northern side of the jet or on the opposite side of the 3C 273 quasar. Let us note, that the spectral analysis of the southern radio lobe was suggested to provide another argument for the sub-relativistic forward velocity of the 3C 273 kpc-scale jet. As mentioned in the previous section, the spectral index of this region is very steep, $\alpha \sim 1.5$, suggesting spectral ageing effects and hence the inferred age of the radiating electrons $\sim 5 \times 10^6$ yrs \citep[for the assumed `typical' lobe parameters;][]{dav85}. Hence, the lobe itself was assumed to be at least that old, implying forward speed of the jet head -- which had to propagate during this time over $\sim 10''$, for a rough estimate -- to be $\beta_{\rm head} \sim 0.01$.

3C 273 was also observed in the past in the search for its hypothetical Mpc-sized radio structure. The low-brightness radio source at $\sim 8.4'$ from the 3C 273 quasar, in the direction of its large-scale jet, was found \citep{rei80,kro83,fol85}. However, relatively flat radio spectrum ($\alpha_{\rm R} \sim 0.85$), as well as the lack of the bridge emission connecting this source with the core of 3C 273, indicate that it is rather a background source, unrelated to the 3C 273 radio structure.

\section{Intermittent jet activity in 3C 273}

The morphology of the large-scale jet in 3C 273 -- its problematically narrow radio cocoon in particular -- resembles in many aspects inner radio lobes of double-double radio galaxies \citep{sch00a,sch00b,sar02,sar03}. Here we propose that in fact this cocoon is a lobe of a new-born jet formed after some period of quiescence following a previous active epoch. If this is the case, then the 3C 273 jet propagates not within the original intergalactic medium of the X-ray gas belonging to the group of nearby galaxies, but within the rarefied plasma formed by the previous jet ejection, i.e. within hypothetical outer, extended radio lobe of 3C 273 source. Then, narrowness of the radio cocoon/inner lobe can be understood in terms of propagation of a ballistic jet \citep{cla91}. Also, the advance velocity of the jet head may be much larger than the typical advance velocity found for the powerful radio sources \citep[which is $< 0.15 \, c$;][]{sch95}, and than the advance velocity estimated for 3C 273 jet from the ram-pressure arguments as presented in the previous section. This allows for the combined Doppler-hiding and time-travel effects to explain the lack of the counter-jet, counter-head and high-brightness part of the counter-lobe in the discussed object, if the jet inclination on the large scales is small (similarly to the pc-scale jet) and the flow velocity (including the head region) is relativistic. We note, that the models ascribing apparent one-sideness of the 3C 273 jet to the relativistic effects solely was discussed before in the literature \citep[e.g.,][]{kun86}. However, intermittent activity of the central engine considered in this paper allows for the first time to put this idea on a firm ground of the standard jet model, giving possible justification for the parameters needed for a Doppler-hiding of the counter-structure in the discussed quasar.

\subsection{Hot-spot region}

We assume that the large-scale jet in 3C 273 is cylindrical, relativistic and dynamically dominated by cold protons \citep[for a justification of the latter assumption see][]{sik00}. Let us note, that all extragalactic large-scale jets possessing optical counterparts are believed to be significantly beamed \citep[see a discussion in][and references therein]{staw04}. Under the above assumptions the jet is highly supersonic with respect to its internal sound speed, and the jet kinetic power is
\begin{equation}
L_{\rm jet} = \pi R_{\rm jet}^2 \, \beta_{\rm jet} c \, \Gamma_{\rm jet}^2 \, u'_{\rm jet} \quad ,
\end{equation}
\noindent
where $R_{\rm jet}$ is the jet cross-section radius, $\Gamma_{\rm jet} \equiv (1 - \beta_{\rm jet}^2)^{-1/2} \gg 1$ is the jet Lorentz factor and
\begin{equation}
u'_{\rm jet} \sim m_{\rm p} \, n'_{\rm p, \, jet} \, c^2
\end{equation}
\noindent
is the jet comoving energy density with proton number density $n'_{\rm p, \, jet}$. In the rest frame of the jet head, the ram-force of the ambient medium is
\begin{equation}
\rho_{\rm amb} \Gamma_{\rm head}^2 \, \left(\beta_{\rm head} c\right)^2 \, A_{\rm head} \quad ,
\end{equation}
\noindent
where $\rho_{\rm amb} = m_{\rm p} \, n_{\rm amb}$ is the density of ambient gas around the jet head, $\Gamma_{\rm head} \equiv (1 - \beta_{\rm head}^2)^{-1/2}$ is the advanced Lorentz factor of the jet head and $A_{\rm head}$ is the cross-section area of a bow-shock at the end of the jet (which can be larger than the cross-section area of the jet itself, $A_{\rm jet} = \pi R_{\rm jet}^2$). Here we neglect pressure of the non-thermal component within the outer lobe, what will be justified below. The ram-force of the ambient medium is balanced by the momentum flux of the jet,
\begin{equation}
m_{\rm p} \, n'_{\rm p, \, jet} \Gamma_{\rm rel}^2 \, \left(\beta_{\rm rel} c\right)^2 \, A_{\rm jet} \quad ,
\end{equation}
\noindent
where $\Gamma_{\rm rel} \equiv (1 - \beta_{\rm rel}^2)^{-1/2}$ is the bulk Lorentz factor of the jet as measured in the rest frame of the head. With relation $\Gamma_{\rm rel} \, \beta_{\rm rel} = \Gamma_{\rm head} \, \Gamma_{\rm jet} \, ( \beta_{\rm jet} - \beta_{\rm head} )$ one can therefore find expression for the advance velocity of the jet head
\begin{equation}
\beta_{\rm head} = \beta_{\rm jet} \, {\sqrt{\eta \, \zeta} \, \Gamma_{\rm jet} \over 1 + \sqrt{\eta \, \zeta} \, \Gamma_{\rm jet}} \quad ,
\end{equation}
\noindent
where
\begin{equation}
\eta \equiv {n'_{\rm p, \, jet} \over n_{\rm amb}} \quad {\rm and} \quad \zeta \equiv {A_{\rm jet} \over A_{\rm head}} \quad .
\end{equation}
\noindent
In the non-relativistic limit with $\beta_{\rm head} \ll \beta_{\rm jet} \ll 1$ expression 5 simplifies to the appropriate equation given by \citet{beg89}. A general formula for $\beta_{\rm head}$ is derived in, e.g., \citet{miz04}.

On the other hand, the brightness asymmetry between the head and the possible counter-head is related to the velocity $\beta_{\rm head}$ and the jet viewing angle $\theta$ by
\begin{equation}
R \equiv {S_{\rm head}(\nu) \over S_{\rm c-head}(\nu)} = \left( {1 + \beta_{\rm head} \, \cos \theta \over 1 - \beta_{\rm head} \, \cos \theta}\right)^{3+\alpha}
\end{equation}
\noindent
as appropriate for a moving source with spectral index $\alpha$ (defined in a way $S(\nu) \propto \nu^{-\alpha}$). Here we assume the counter-head to be identical with the head in physical properties. We also consider the simplest case in which these properties did not change significantly during the recent phase of the source evolution, i.e. between the time at which the receding counter-head and the approaching head emitted the observed radiation. Hence, for the known parameter $R$ one can find the required advanced velocity of the jet head as a function of the jet viewing angle,
\begin{equation}
\beta_{\rm head} = {R^{1/(3+\alpha)} - 1 \over R^{1/(3+\alpha)} + 1} \, { 1 \over \cos \theta} \quad ,
\end{equation}
\noindent
and then, using equation 5, the required density contrast between the jet and the surrounding medium as
\begin{equation}
\eta = {\beta_{\rm head}^2 \over \zeta \, \Gamma_{\rm jet}^2 \, \left(\beta_{\rm jet} - \beta_{\rm head}\right)^2} \quad .
\end{equation}
\noindent
One may note, that a more sophisticated (although inevitably arbitrary) model including a decrease in the luminosity of the expanding head region during its evolution, would increase the required value of the head velocity $\beta_{\rm head}$, and therefore of the parameter $\eta$ \citep[see][]{mil04}.

Now let us discuss what is the required value of $\eta$ for the 3C 273 jet, if one assumes that lack of the counter-head in this source is due to Doppler-hiding effects alone. Firstly, for the required radio brightness asymmetry $R = 10^4$ and the radio spectral index $\alpha = 0.8$ one can find from the expression 8 that the advance velocity of the jet head is $\beta_{\rm head} = 0.837 / \cos \theta$. This velocity, as well as the appropriate Lorentz and Doppler factors of the head, are shown on figure 1 for the jet viewing angles $\theta = 10^{\circ} - 30^{\circ}$. Note that they are the lower limits since the head/counter-head radio brightness asymmetry can be in reality higher than $10^4$. Note also that for $\theta \sim 10^{\circ} - 20^{\circ}$ the head Doppler factor $\delta_{\rm head} = [\Gamma_{\rm head} \, (1 - \beta_{\rm head} \, \cos \theta)]^{-1} \sim 3$ is high enough to explain the change in the jet polarization within the head region by the compression and relativistic aberration effects, without invoking large bending of the flow between pc- and kpc-scales \citep[see section 2.3;][]{fla85}. Next, in order to find $\eta$, we make an additional assumption that $\zeta < 1$, what is justified in a framework of our interpretation of the radio cocoon as the inner lobe. In such a case the jet radius can be inferred from optical observations\footnote{Throughout the paper we assume $\Omega_{\rm M} = 0.3$, $\Omega_{\Lambda} = 0.7$ and $H = 70$ km s$^{-1}$ Mpc$^{-1}$ cosmology, leading to the conversion scale $2.7$ kpc$/''$ at the 3C 273 redshift $z=0.158$.} as $R_{\rm jet} = 1.0$ kpc, while the bow-shock radius at the end of the jet can be identified with the radius of the radio structure, $R_{\rm head} = 1.4$ kpc, leading to $\zeta = (R_{\rm jet} / R_{\rm head})^2 \sim 0.5$. Parameter $\eta$ (basically its lower limits) as a function of the jet bulk Lorentz factor $\Gamma_{\rm jet} = 5 - 10$ is shown on figure 2 for different jet viewing angles $\theta = 10^{\circ}, \, 20^{\circ}$ and $30^{\circ}$, parameters $R=10^4$ and $\zeta = 0.5$. For $\theta \sim 10^{\circ} - 20^{\circ}$ the required density contrast between the jet and the ambient medium is roughly $\eta \sim 1 - 10$. This means that the jet is required to be over-dense with respect to the ambient medium, what is in disagreement with the standard model of powerful extragalactic jets, with possible exception for the inner beams in double-double radio galaxies.

Powerful radio sources at low redshifts ($z \sim 0.1$) are typically located in isolated elliptical galaxies or, like 3C 273, in poor groups of galaxies. The density of the intergalactic medium in such environments, i.e. density of hot intergalactic gas, can be found from its thermal X-ray emission. Observations indicate, that the gas density within typical poor group of galaxies at distance $r$ from the group center can be approximated by
\begin{equation}
n_{\rm igm} = n_{0} \, \left( {r \over a_{0}} \right)^{- \beta} \quad ,
\end{equation}
\noindent
with $r > a_{0} = 10$ kpc, $\beta = 1.5$ and $n_{0} = 10^{-2}$ cm$^{-3}$ \citep{gar91,mul98,blu99,wil99,kai00}. Hence, one can estimate the expected number density of the original (unperturbed) intergalactic gas surrounding 3C 273 large-scale jet, by substituting into equation 10 for $r$ the physical distance of the jet head from the active center, $d_{\rm head} = d_{\rm proj} / \sin \theta$. For the projected separations of the H2 component from the 3C 273 quasar $\sim 21.4''$, i.e. for $d_{\rm proj} \sim 60$ kpc, one gets the expected number density of the intergalactic medium ranging from $n_{\rm igm} \sim 0.5 \times 10^{-4}$ cm$^{-3}$ for $\theta = 10^{\circ}$, up to $n_{\rm igm} \sim 2.4 \times 10^{-4}$ cm$^{-3}$ for $\theta = 30^{\circ}$. Below, we take the average value $n_{\rm igm} \sim 10^{-4}$ cm$^{-3}$ in agreement with estimations by \citet{con81,con93}. However, the real density of the ambient medium around 3C 273 jet can be much lower than this, because of previous jet active periods. Therefore, we write
\begin{equation}
n_{\rm amb} = \kappa \times n_{\rm igm} \quad .
\end{equation}
\noindent
In the standard evolution model of powerful radio sources \citep{beg89} the radio lobes are expected to be empty from the thermal intergalactic matter, and hence the number density of the lobe medium to be solely due to the particles deposited within the lobes by the pair of jets. As the jet matter, before entering the lobes, is shocked within the hot-spot region, the gas within the lobes is expected to be relativistic, i.e. its rest energy density should be negligible when compared to its pressure. However, observations of double-double radio galaxies \citep{sch00a} indicate the mass density within the outer lobes in excess over the one expected in such a standard picture, suggesting that the old radio lobes have to be enriched by some additional thermal matter. \citet{kai00} proposed, that this additional material is warm and dense clouds of gas embedded originally in the hot intergalactic medium, which are destroyed and dispersed over the old lobe volume (with the time scale $\sim 10^7$ yrs) during its evolution. In the framework of this model one expects $\kappa \sim 0.01 - 0.001$. In such a case, even with $\kappa \, n_{\rm igm} = 10^{-7}$ cm$^{-3}$ the rest-mass energy density of the outer lobe medium, $\rho_{\rm amb} \, c^2 = m_{\rm p} \, \kappa \, n_{\rm igm} \, c^2 \sim 1.5 \times 10^{-10}$ erg cm$^{-3}$, is still much larger than pressure of the non-thermal component within the outer lobe, which is estimated in the next section as $p_{\rm lobe} \ll 10^{-10}$ erg cm$^{-3}$ (equation 19). This justifies omission of the latter one in formula 3.

The proton number density of the jet can be found from equations 1-2 as
\begin{equation}
n'_{\rm p, \, jet} \sim {L_{\rm jet} \over \pi R_{\rm jet}^2 \, \beta_{\rm jet} \, c^3 \, \Gamma_{\rm jet}^2 \, m_{\rm p}} \quad .
\end{equation}
\noindent
Hence, the density contrast between the jet and the ambient medium can be written using equations 11 and 12 as
\begin{equation}
\xi = {n'_{\rm p, \, jet} \over n_{\rm amb}} \sim {0.74 \over \kappa \, \Gamma_{\rm jet}^2} \, {L_{\rm jet, \, 47} \over n_{\rm igm, \, -4}}
\end{equation}
\noindent
for $R_{\rm jet} \sim 1$ kpc, $\beta_{\rm jet} \sim 1$, $L_{\rm jet, \, 47} \equiv L_{\rm jet} / 10^{47}$ erg s$^{-1}$ and $n_{\rm igm, \, -4} \equiv n_{\rm igm} / 10^{-4}$ cm$^{-3}$, where we use letter $\xi$ in order to distinguish this quantity from the density contrast $\eta$ required by the head/counter-head brightness asymmetry (equations 8 and 9). Parameter $\xi$ as a function of the jet Lorentz factor $\Gamma_{\rm jet} = 5 - 10$ is presented on figure 2, for $L_{\rm jet, \, 47} = 1$ \citep[expected in the case of a powerful quasar jet;][]{raw91,wil99}, $n_{\rm igm, \, -4} = 1$ and $\kappa = 0.001, \, 0.01$ and $1$. As illustrated, for jet viewing angles $\theta = 10^{\circ} - 20^{\circ}$ and parameter $\kappa = 0.001 - 0.01$ both values of $\eta$ and $\xi$ can be reconciled.

\subsection{Jet, counterjet and inner lobes}

The head of the jet moving with advance velocity $\beta_{\rm head}$ travels the projected distance $d_{\rm proj}$ during the time
\begin{equation}
t = {d_{\rm proj} \over c \, \beta_{\rm app}} \quad ,
\end{equation}
\noindent
where
\begin{equation}
\beta_{\rm app} = {\beta_{\rm head} \, \sin \theta \over 1 - \beta_{\rm head} \, \cos \theta} \quad .
\end{equation}
\noindent
Hence, one can find the age of the large-scale jet in 3C 273 by substituting into equation 14 the head velocity $\beta_{\rm head} = 0.837 / \cos \theta$ (equation 8 for $R = 10^4$ and $\alpha = 0.8$) and separation of the jet head from the core $d_{\rm proj} = 60$ kpc, to obtain $t_{\rm jet} \sim 3.8 \times 10^4 \, \cot \theta$ yrs. For $\theta = 10^{\circ}$ it is $t_{\rm jet} \sim 2.1 \times 10^5$ yrs, and for $\theta = 30^{\circ}$ one obtains $t_{\rm jet} \sim 0.7 \times 10^5$ yrs. Thus, we take the age of the large-scale jet in 3C 273 to be roughly $t_{\rm jet} \sim 10^5$ yrs. As the head/counter-head brightness asymmetry gives only a lower limit on $\beta_{\rm head}$, value $10^5$ yrs should be considered as an upper limit for the time since formation of the considered structure.

The ratio of the projected separation of the jet head from the active center to the projected separation of the counter-head is
\begin{equation}
Q \equiv {d_{\rm proj} \over d_{\rm c-proj}} = {1 + \beta_{\rm head} \, \cos \theta \over 1 - \beta_{\rm head} \, \cos \theta} = R^{1 / (3+\alpha)} \quad .
\end{equation}
\noindent
For $R = 10^4$ and $\alpha = 0.8$ one gets $Q = 11.3$. That is, the projected distance of the counter-head from the quasar core is expected to be $d_{\rm c-proj} =  d_{\rm proj} / Q = 1.9'' \sim 5$ kpc. Now, the optical (and X-ray) jet is visible starting from $\sim 12''$ from the core. It is not clear, if this gap between the active center and the first appearance of the optical structure is due to Doppler-hiding of the jet emission at distances $< 12''$, or if it is due to intermittent/modulated jet activity leading to the formation of the `partial' jet. Note, that the continuous radio emission in this region can be connected with the `false jet structure' obtained in the numerical simulation of the restarting jets by \citet{cla91}, i.e. with the cocoon material collapsing on the abandoned jet channel. One may also note, that the modulation in the jet kinetic power can involve solely modulation of the particle flux density, and not the jet bulk velocity. If one agrees that the gap in the optical (and X-ray) emission of the 3C 273 jet is indeed due to modulated activity of the central engine, i.e. that the projected length of the considered large-scale jet is $l_{\rm jet} \sim 10''$, the expected length of the counterjet is
\begin{equation}
l_{\rm c-jet} = l_{\rm jet} \,  {1 - \beta_{\rm jet} \, \cos \theta \over 1 + \beta_{\rm jet} \, \cos \theta} \quad .
\end{equation}
\noindent
This parameter, as a function of the jet viewing angle $\theta = 10^{\circ} - 30^{\circ}$, is shown on figure 3 for $\Gamma_{\rm jet} = 5$ and 10. For comparison, the upper limit for $d_{\rm c-proj}$ is also presented. Therefore, the Doppler and time-travel effects can easily explain lack of the counter-jet at the opposite side of the 3C 273 quasar. We propose, that also the apparent lack of the high surface brightness part of the counter-lobe (analogous to the radio cocoon enveloping $12'' - 23''$ jet) can be explain in a similar way. Note, that for a relativistic velocity of the jet head, the jet matter turned-back at the hot-spot region may be still relativistic in the observer frame. This issue cannot be discussed in more details in the present paper, but we believe that it constitutes an interesting problem for the numerical simulations referring to propagation of \emph{relativistic} and \emph{restarting} jets \citep[see also in this context the case of radio galaxy 3C 219 discussed in][and in the last section of this paper]{bri86,cla92}.

Let us now comment on the high frequency, optical-to-X-ray emission of the large-scale jet in 3C 273. Highly relativistic bulk velocities and small viewing angle considered here are consistent with interpretation of the X-ray emission of this jet as being due to comptonization of the CMB photon field \citep{sam01}. However, this model faces several difficulties in explaining some properties of the powerful quasar jets, and the 3C 273 jet knotty structure in particular \citep[see][]{staw04,sta04}. The jet knotty structure is usually interpreted in terms of strong and extended shock waves residing within the knot regions. Note however, that magnetic field parallel to the jet axis, as inferred from \emph{both} radio and optical observations of 3C 273, disagree with internal shocks model, while frequency-independent knot profiles -- especially distinct in 3C 273 jet -- are not expected in a framework of the stationary shock scenario. Let us also mention, that standard shocks-in-knots models cannot explain easily neither the spectral character of the optical emission in this object, as emphasized by \cite{jes01,jes02}. \citet{sta04} proposed, that the knots observed in the large-scale quasar jets represent not single extended shock waves, but moving jet portions with the excess of kinetic power due to modulated activity of the jet engine. When applied to the 3C 273 jet as considered in this paper, this proposition would mean that the $\sim 10^5$-year-old jet is modulated in its kinetic power on the characteristic time scale $\sim 10^4$ yrs.

\subsection{Outer structures}

Obviously, the presented model of the 3C 273 large-scale jet, being an analogue of the inner radio structures observed in double-double radio galaxies, assumes the presence of some extended, old radio lobe surrounding the source. It is, however, not clear if such a low-brightness old radio structure, even if present, can be detected (also because of extremely high brightness of the 3C 273 quasar core). It is unknown when such a structure was eventually formed, and therefore what should be its size and radio luminosity. It is not even certain how many jet active periods 3C 273 has undergone in the past, or what was their duration and kinetic power. One can speculate if the southern extension of the 3C 273 radio jet (section 2.4) is in fact some fragmentary remnant of the older lobe. However, the smooth continuation of the radiating plasma between the radio jet (cocoon) and this region questions, to some extent, such a possibility.

For the small jet viewing angle considered here, the eventual outer lobe would be seen in projection, therefore displaying a more spherical than elongated spheroidal shape. The radius of such a structure, $D$, would therefore depend on the sideway expansion of the radio lobe, controlled by the non-thermal outer lobe pressure. This pressure can be roughly estimated by dividing the total energy transported by the jets during previous active epoch, $W \equiv \int \, L(t) \, dt$, by the volume of the outer lobe, $V_{\rm lobe}$,
\begin{equation}
p_{\rm lobe} \sim \left(\hat{\gamma}_{\rm lobe} - 1\right) \, {2 \, W \over V_{\rm lobe}} \quad ,
\end{equation}
\noindent
where the ratio of the specific heats is $\hat{\gamma}_{\rm lobe} = 4/3$ as appropriate for the non-thermal lobe plasma. For the outer lobe we assume spheroidal geometry and take $V_{\rm lobe} = 4 \pi \, D^3$ with the axis ratio $= 3$, typical for the powerful radio sources (and also for outer lobes in double-double radio galaxies). As mentioned above, due to projection effects the observed radius of the other lobe is expected to be roughly equal to $D$, while the physical extent of the whole outer structure is $\sim 6 \, D$. For the previous jet activity we assume kinetic power and the lifetime similar to the present jet, and take $W \sim L_{\rm jet} \, t_{\rm jet}$. Thus, the non-thermal pressure within the outer lobe is
\begin{equation}
p_{\rm lobe} \sim 5.7 \times 10^{-13} \, W_{59.5} \, D_{2}^{-3} \quad {\rm dyn \, cm^{-2}} \quad ,
\end{equation}
\noindent
where $W_{59.5} \equiv W / 10^{59.5} \, {\rm erg} \sim 1$ refers to the case $L_{\rm jet} \sim 10^{47}$ erg s$^{-1}$ and $t_{\rm jet} \sim 10^5$ yrs, and $D_{2} \equiv D / 10^2$ kpc is expected to be $\sim 1$ in analogy to double-double radio galaxies. On the other hand, pressure of the thermal intragroup gas at distance $r$ from the group center is
\begin{equation}
p_{\rm igm} \sim n_{\rm igm} \, k T = n_{0} \, k T \, \left( {r \over a_{0}} \right)^{- \beta} \quad ,
\end{equation}
\noindent
where $T \sim 10^7$ K is the temperature of the hot ambient gas, and we take parameters $n_0$, $a_0$ and $\beta$ as in the previous section (see equation 10). Hence, the ratio of the pressures within the outer lobe and the ambient medium at the lobe edges is
\begin{equation}
\chi \equiv {p_{\rm lobe} \over p_{\rm igm}} \sim 1.3 \, W_{59.5} \, D_{2}^{-1.5} \quad .
\end{equation}
\noindent
For $\chi > 1$ the outer lobe expansion is pressure-driven, and the ram-pressure condition at the contact discontinuity gives the sideway expansion velocity of the outer lobe
\begin{equation}
v_{\rm lobe} \sim \left({p_{\rm lobe} \over \rho_{\rm igm}}\right)^{1/2}
\end{equation}
\noindent
provided that the expansion is supersonic with respect to the external medium \citep[e.g.,][]{beg89}. For such a supersonic expansion, the shock between the outer lobe and the intragroup matter can form. In such a case, one may hope to observe the contact discontinuity between the outer lobe and the unperturbed intergalactic (intragroup) medium due to optical line emission excited near the shock front \citep[see][]{bic96}, or the enhanced X-ray thermal emission due to hot ambient gas compressed at such a shock (i.e., the sideway bow-shock of the relict cocoon). The sound speed in the ambient medium is
\begin{equation}
c_{\rm s, \, igm} = \left(\hat{\gamma}_{\rm igm} \, {p_{\rm igm} \over \rho_{\rm igm}}\right)^{1/2} = \left({5 \over 3} \, {k T \over m_{\rm p}}\right)^{1/2} \sim 3.7 \times 10^7 \quad {\rm cm \, s^{-1}}
\end{equation}
\noindent
for $T = 10^7$ K and $\hat{\gamma}_{\rm igm} = 5/3$ as appropriate for the considered thermal matter. Therefore, one can write
\begin{equation}
\beta_{\rm s, \, lobe} \equiv {v_{\rm lobe} \over c_{\rm s, \, igm}} \sim 0.8 \, \chi^{1/2} \quad .
\end{equation}
\noindent
Parameters $\chi$ and $\beta_{\rm s, \, lobe}$ as functions of the outer lobe radius $D = 50 - 500$ kpc are shown on figure 4 for $W_{59.5} = 1$ and $10$. In the less optimistic scenario, i.e. for $W_{59.5} = 1$, the bow-shock features could be eventually present only if the age of the outer structure is low enough to allow for $D_{2} < 1$. However, if the previous jet activity was longer, or equivalently the previous jet was more energetic leading to $W_{59.5} = 10$, the shock features could be eventually observed even up to $D_{2} \sim 4$.

Now let us give a very rough estimate of the expected radio flux from the outer lobe. The total synchrotron luminosity is
\begin{equation}
L_{\rm syn} = {4 \over 3} \, {c \sigma_{\rm T} \over m_{\rm e} c^2} \, {\langle \gamma^2 \rangle \over \langle \gamma \rangle} \, u_{\rm B, \, lobe} \, V_{\rm lobe} \, u_{\rm e, \, lobe} \quad ,
\end{equation}
\noindent
where $u_{\rm e, \, lobe}$ is the electron energy density within the outer lobe, $\langle \gamma^2 \rangle$ and $\langle \gamma \rangle$ are the respective moments of the electron energy distribution, and $u_{\rm B, \, lobe} = B^2 / 8 \pi$ is the outer lobe magnetic field energy density. In accordance with the anticipated model of proton-dominated jet, we assume that the fraction of the total lobe energy density deposited in relativistic electrons, as well as in the magnetic field, is smaller than unity,
\begin{equation}
\varepsilon_{\rm e, \, lobe} \equiv {u_{\rm e, \, lobe} \over u_{\rm lobe}} < 1 \quad , \quad \varepsilon_{\rm B, \, lobe} \equiv {u_{\rm B, \, lobe} \over u_{\rm lobe}} < 1 \quad ,
\end{equation}
\noindent
where the total energy density of the outer lobe is $u_{\rm lobe} = 3 \, p_{\rm lobe} \sim 2 \, W / V_{\rm lobe}$. For the electron energy distribution we assume the `evolved' power-law form $\propto \gamma^{-3}$, with low and high energy cut-offs $\gamma_{\rm max} \gg \gamma_{\rm min}$. Thus, $\langle \gamma^2 \rangle / \langle \gamma \rangle \sim \gamma_{\rm min} \, \ln(\gamma_{\rm max} / \gamma_{min})$, and the synchrotron luminosity is
\begin{equation}
L_{\nu} = k \, \nu^{-1} \quad ,
\end{equation}
\noindent
where coefficient $k$ can be found as
\begin{equation}
k = {L_{\rm syn} \over \ln(\nu_{\rm max} / \nu_{\rm min})} \sim {\sigma_{\rm T} \over 6 \pi \, m_{\rm e} c} \, \gamma_{\rm min} \, B^2 \, W \, \varepsilon_{\rm e, \, lobe} \quad .
\end{equation}
\noindent
The flux density of the synchrotron emission is then
\begin{equation}
S_{\nu} = {L_{\nu} \over 4 \pi \, d_{\rm L}^2} \sim 3 \times 10^{-18} \, B_{\mu {\rm G}}^2 \, \gamma_{\rm min} \, W_{59.5} \, \varepsilon_{\rm e, \, lobe} \, \nu^{-1} \quad {\rm erg \, s^{-1} \, cm^{-2} \, Hz^{-1}} \quad ,
\end{equation}
\noindent
where $d_{\rm L} = 1025$ Mpc is the luminosity distance to 3C 273, and $B_{\mu {\rm G}} \equiv B / \mu {\rm G}$. For the model parameters $\gamma_{\rm min} \sim 10$, $B_{\mu {\rm G}} \sim 3$, $W_{59.5} \sim 1$ and $\varepsilon_{\rm e, \, lobe} \sim 0.2$ (referring to the energy equipartition between the magnetic field and the radiating electrons for the outer lobe radius $D_{2} \sim 1$), one gets the $1.4$ GHz flux $\sim 4$ mJy, which is $\sim 10^{-4}$ of the total 3C 273 flux at this frequency. Obviously, the above estimates are illustrative only. However, we believe that any more sophisticated approach would be at the present stage as arbitrary as the present one. That is because the number of the unknown old lobe parameters would increase significantly in such a case. In particular, using `more realistic' spectrum for the old lobe electrons would not necessarily result in `more realistic' prediction about the expected radio flux. We note many controversies regarding the complicated issue of modeling the electron spectral evolution in the lobes of powerful radio sources \citep[e.g.,][]{blu00,ka00,man02}.

The relativistic electrons deposited within the outer lobe during the eventual previous jet active period can be pronounced not only at radio frequencies, but also at X-ray photon energies due to inverse-Compton scattering on the CMB photon field \citep{har79} or on the far-infrared radiation produced by the 3C 273 quasar core \citep{bru97}. Such a non-thermal X-ray emission was already detected from the number of lobes in FR II radio galaxies and quasars \citep[see, e.g.,][and references therein]{bru02}. In this context, it is tempting to speculate if the X-ray halo detected by ROSAT around 3C 273 \citep{roe00} is not due to the intergalactic hot gas, but consists of two different components: the thermal one originating in the interstellar medium of the 3C 273 host galaxy \citep[see][]{mul98} and the extended, non-thermal high-energy tail due to relativistic electrons of the outer lobes. However, we estimate the total X-ray luminosity of the outer lobe due to comptonization of the CMB radiation (in analogy to the radio luminosity estimated above) as $\sim 10^{40}$ erg s$^{-1}$, which seems to be insufficient for a clear detection.

\section{Discussion and conclusions}

In this paper we propose that the large-scale jet in 3C 273 is an analogue of the inner lobes observed in double-double radio galaxies. This model can explain the narrowness of the radio cocoon in the considered source, as well as the lack of the counterjet, counter-lobe and the counter-head on the opposite side of the active center, if the jet inclination on the large scales is small, $\theta \sim 10^{\circ} - 20^{\circ}$, the jet is matter-dominated and relativistic up to its head, $\Gamma_{\rm jet} = 5 - 10$, and if the original intragroup medium surrounding 3C 273 is rarefied by previous jet activity by a factor of $100 - 1000$. If this is the case, then 3C 273 constitutes a quasar-counterpart to a few known double-double sources, which are inclined at much larger angles to the line of sight, $\sim 90^{\circ}$ \citep[see][]{sch00a,sch00b,sar02,sar03}. Let us mention that the intermittent case between 3C 273 and other `classical' double-doubles can be radio galaxy 3C 219. The jet in this source shares many morphological similarities to the 3C 273 jet, and is considered by some authors as the example of the restarting (`partial') beam \citep{bri86,cla92}. The overall radio morphology of 3C 219 was recently interpreted as a double-double one, but only observed at intermittent jet viewing angles \citep{sch00a}. Interestingly, on the opposite side of the 3C 219 active center a single radio knot was detected and subsequently identified with the counter-head of the restarting counter-jet \citep{cla92}. The ratio of the projected distances of the head and the counter-head from the 3C 219 core is $\sim 3.6$, which can be compared with the appropriate value for double-double radio galaxies, $\sim 1$, and for 3C 273, as estimated in this paper, $> 11$.

Besides the DDRGs, decaying activity of the extragalactic radio sources manifests in formation of `radio relics' found in rich clusters of galaxies. However, we believe that the analogy between 3C 273 and the DDRGs is more appropriate than the analogy between this object and cluster radio sources, for the following reasons. First, we note unusual narrowness of the large-scale radio jet/cocoon in 3C 273, which resembles morphology of the inner lobes of DDRGs. Second, 3C 273 is located in poor group of galaxies (similarly to the DDRGs) and not in rich cluster like radio relics considered by, e.g., \citet{dey03}. As the evolution of radio lobes depends crucially on the properties of the surrounding medium, we believe that the model by \citet{kai00} regarding extended outer lobes of DDRGs is more appropriate to adopt in this paper than the models regarding radio relics observed in clusters \citep[see in this context][]{kom94,kai02}. Also, the DDRGs -- exhibiting in addition `pure' FR II morphology -- constitutes the \emph{direct} proof for the intermittent/recurrent jet activity in powerful radio sources. Ghost radio cavities found in galaxy clusters are widely interpreted as the remnants of ceased radio activity, but there is no observed case for the new-born FR II jet propagating within the cluster radio relic \citep[but see][]{jon01,hei02,ven04}.

We estimated the age of the large-scale jet in 3C 273 as $\sim 10^5$ yrs. This would be then the time-scale of the jet active period. Unfortunately, the time-scale of the quiescence epoch before the present activity cannot be evaluated at the moment, as we did not detect its outer lobes until now (but see below). In analogy to the double-double radio galaxies one can only suggest that it is $\geq 10^7$ yrs. We also propose, that knotty morphology of the discussed jet indicate modulation in the jet kinetic power on the time scale $\sim 10^4$ yrs \citep[see][]{sta04}. This, together with the VLBI jet variability time scale of the order of years, indicates that the jet activity in 3C 273, and possibly in other similar sources, is variable/modulated/intermittent over many decades of the source lifetime. This may have several potential consequences for the jet formation models \citep[see in this context also][]{rey97,sie97,mar03}. We note, that recent coronographic observations of quasar 3C 273 with the Advanced Camera for Survays \citep{ma03} revealed complicated morphology of the host galaxy, with the complex structure of filaments and clumpy regions, spiral-shaped plume, dust lanes, etc. It was suggested, that the possible explanation for such a complexity could be a relatively recent merger of 3C 273 with a massive companion galaxy. It is then possible, that such an eventual merger event started the epoch of the \emph{recurrent} jet activity in the considered source, resulting in formation of the extended outer lobes some $10^7 - 10^8$ yrs ago, and next of the present $10^5$-year-old jet \citep[see in this context][]{mer02,gop03}.

During the review process for this paper the author became aware of the new, unpublished $1.4$ GHz map of 3C 273 at $4''$ FWHM resolution with the dynamic range $30,000$ \citep{per04}. It shows for the very first time some faint extended radio emission surrounding the 3C 273 quasar, enveloping its large-scale jet and also present at the counter-jet side. Interestingly, this extended radio structure seems to be elongated in the direction of the large-scale jet, as expected in the case of double-double radio morphology. Hence, in the near future, with the new data published, the proposed model can be verified and enriched by a more detailed analysis of the old lobe parameters.

\acknowledgments
I am grateful to Micha\l \, Ostrowski and Marek Sikora for the valuable discussions. I also acknowledge very useful comments and suggestions of the referee, Sebastian Jester. I thank Rick Perley for providing me the new $1.4$ GHz map of 3C 273. The present work was supported by the grant PBZ-KBN-054/P03/2001.

\begin{figure}
\plotone{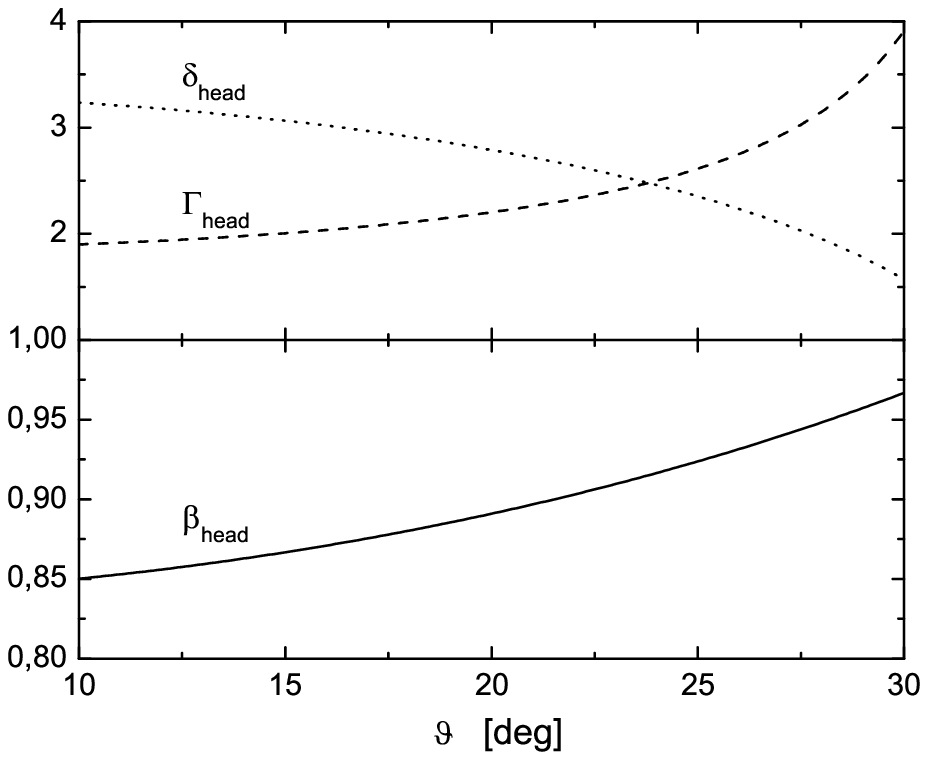}
\caption{ The advance velocity $\beta_{\rm head}$ (solid line), the Lorentz factor $\Gamma_{\rm head}$ (dashed line) and the Doppler factor $\delta_{\rm head}$ (dotted line) of the jet head, as functions of the jet viewing angle $\theta$, for the head/counter-head radio brightness asymmetry $R=10^4$ and the radio spectral index $\alpha = 0.8$. \label{f1}}
\end{figure}

\begin{figure}
\plotone{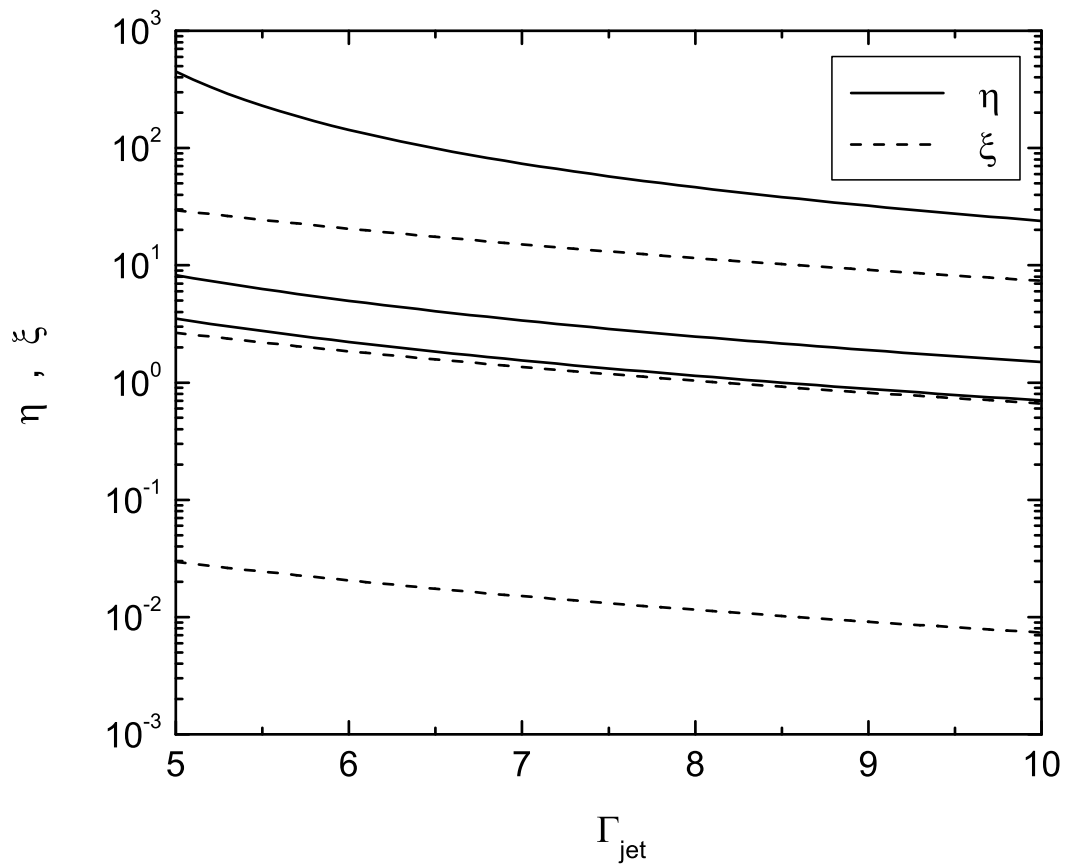}
\caption{ The solid lines represent parameter $\eta$ as a function of the jet bulk Lorentz factor $\Gamma_{\rm jet}$, for $R=10^4$, $\zeta = 0.5$ and jet viewing angles $\theta = 30^{\circ}, \, 20^{\circ}$ and $10^{\circ}$ (from the top to the bottom of the plot, respectively). The dotted lines represent parameter $\xi$ for $L_{\rm jet, \, 47} = 1$, $n_{\rm igm, \, -4} = 1$ and $\kappa = 0.001, \, 0.01$ and $1$ (from the top to the bottom of the plot, respectively). \label{f2}}
\end{figure}

\begin{figure}
\plotone{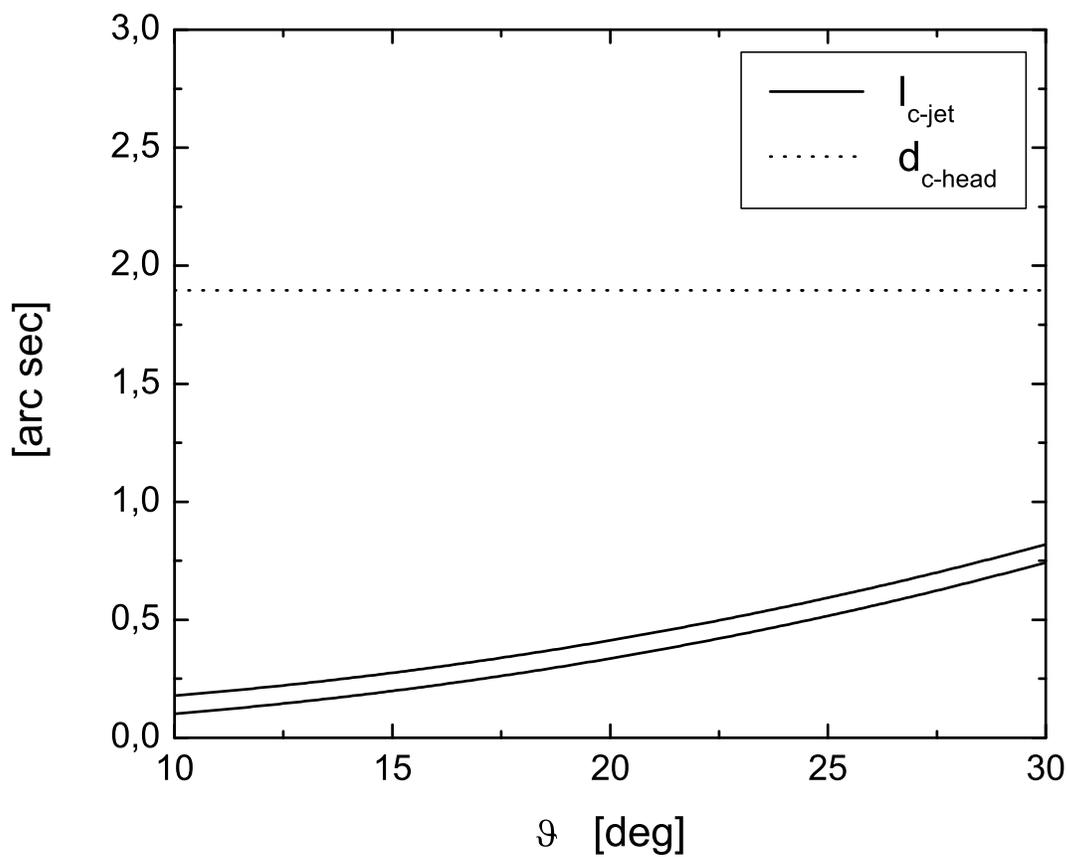}
\caption{ The solid lines represent the expected projection length of the counterjet, $l_{\rm c-jet}$, as a function of the jet viewing angle $\theta$, for the jet bulk Lorentz factor $\Gamma_{\rm jet} = 5$ and $10$ (upper and lower curves, respectively). The dotted line corresponds to the expected separation of the counter-head from the quasar core, $d_{\rm c-head}$, for the parameter $R = 10^4$. \label{f3}}
\end{figure}

\begin{figure}
\plotone{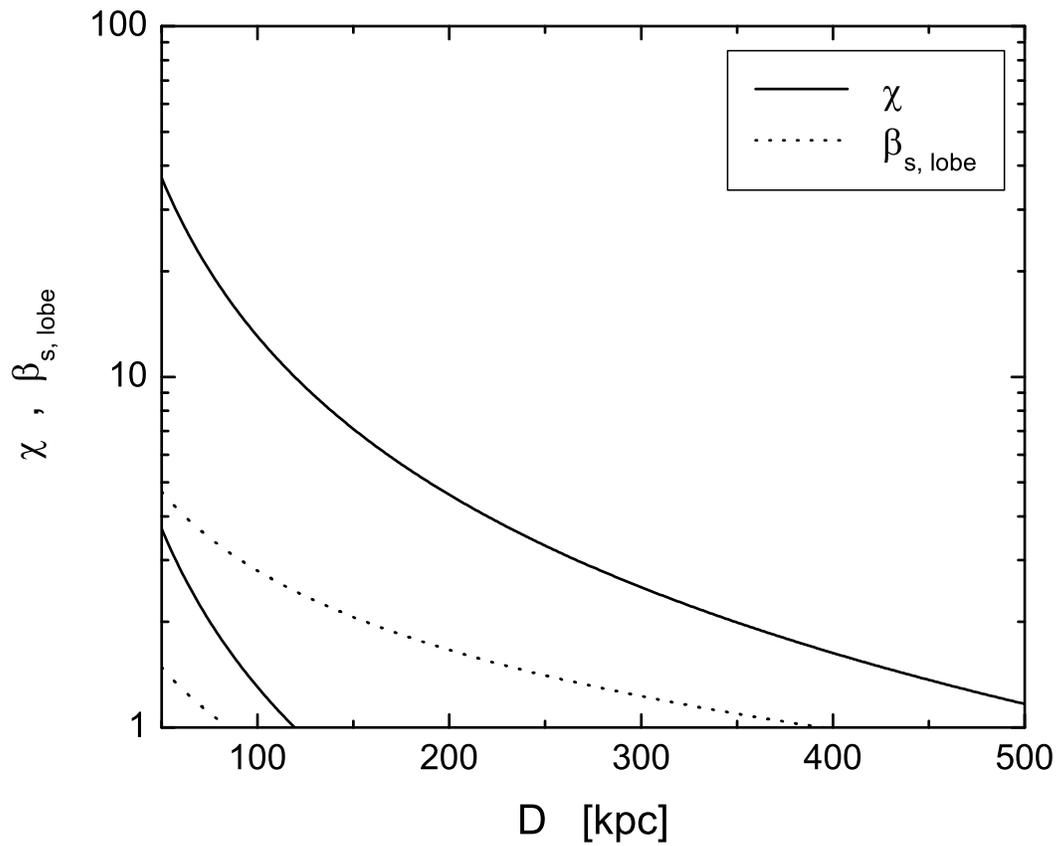}
\caption{ The solid lines represent parameter $\chi$ as a function of the outer lobe radius $D$, for $W_{59.5} = 1$ and $10$ (the lower and upper lines, respectively). The dotted lines represent the appropriate parameter $\beta_{\rm s, \, lobe}$. \label{f4}}
\end{figure}

\end{document}